\documentclass[a4paper]{llncs}

\usepackage{amsmath, amssymb}
\usepackage{inputenc, graphicx, verbatim,color, enumerate}

\usepackage{float}

\usepackage{tikz,pgfplots}
\usepackage{url}
\usepackage{xcolor}

\newcommand{\F}{\mathbb{F}}

\renewcommand{\P}{\mathbb{P}}

\newcommand{\UU}{\mathcal{U}}
\newcommand{\FF}{\mathcal{F}}
\newcommand{\ZZ}{\mathcal{Z}}

\newcommand{\cyc}{\mathrm{cyc}}
\newcommand{\cl}{\mathrm{cl}}
\newcommand{\ie}{\emph{i.e.}}

\newcommand{\rmatroid}{M = (\rho,E)}

\title{On Binary Matroid Minors and Applications to Data Storage over Small Fields}

\author{Matthias Grezet \and Ragnar Freij-Hollanti \and Thomas Westerb\"ack \and Camilla Hollanti}

\institute{Department of Mathematics and Systems Analysis, Aalto University, Finland\\
\email{firstname.lastname@aalto.fi}}

\begin{document}

\maketitle

\abstract{Locally repairable codes for distributed storage systems have gained a lot of interest recently, and various constructions can be  found in the literature. However, most of the constructions result in either large field sizes and hence too high computational complexity for practical implementation, or in low rates translating into waste of the available storage space. 
	
In this paper we address this issue by developing theory towards code existence and design over a given field. This is done via exploiting recently established connections between linear locally repairable codes and matroids, and using matroid-theoretic characterisations of linearity over small fields. In particular, nonexistence can be shown by finding certain forbidden uniform minors within the lattice of cyclic flats. 
It is shown that the lattice of cyclic flats of binary matroids have additional structure that significantly restricts the possible locality properties of $\F_2$-linear storage codes.
Moreover, a collection of criteria for detecting uniform minors from the lattice of cyclic flats of a given matroid is given, which is interesting in its own right. 
}


\section{Introduction}
The need for large-scale data storage is continuously increasing. Within the past few years, \emph{distributed storage systems} (DSSs) have revolutionised our traditional ways of storing, securing, and accessing  data. Storage node failure is a frequent obstacle, making repair efficiency an important objective. Network coding techniques for DSSs were considered in \cite{dimakis10}, characterising a storage space--repair bandwidth tradeoff.  

A bottle-neck for repair efficiency, measured by the notion of \emph{locality} \cite{papailiopoulos12}, is the number of contacted nodes needed for repair. To this end, our motivation in this paper comes from \emph{locally repairable codes} (LRCs), which are, informally speaking, storage systems where a  small number of failing nodes can be recovered by boundedly many other (close-by) nodes. Repair-efficient LRCs are already implemented on HDFS-Xorbas used by Facebook \cite{Facebook} and Windows Azure storage \cite{Windows}. Here, the field size is not yet a huge concern, as the coding is done over a small number ($<20$) of nodes. Nevertheless, if we wish for more flexibility in terms of the code parameters, the field size quickly becomes a critical issue. Hence, a question arises as to how and when can we maintain a small field size, regardless of the number of the storage nodes. Some explicit constructions of LRCs over small fields can be found in the literature, \emph{e.g.}, \cite{cadambe15,huang16,ernvall16,silberstein15,tamo16}.

Let us denote by $(n,k,d,r,\delta)$, respectively, the code length, dimension, global minimum distance, locality, and local minimum distance. In terms of a storage system employing an $(n,k,d,r,\delta)$-LRC, this means that we encode  $k$ information symbols into $n$ code symbols that are then stored on $n$ storage nodes, and can globally tolerate $d-1$ node failures while still being able to repair by contacting $k$ nodes. Locally, if we lose at most $\delta-1$ nodes ($\delta\leq d)$, we can repair those by contacting at most $r<k$ close-by nodes. 

It was shown in \cite{tamo13} that the $(r,\delta = 2)$-locality of a linear LRC is a matroid invariant. The connection between matroid theory and linear LRCs was examined in more detail in \cite{westerback15}. In addition, the parameters $(n,k,d,r,\delta)$ for linear LRCs were generalised to matroids, and new results for both matroids and linear LRCs were given therein. 


In this paper, we develop theory towards matroids that are representable over the binary field or some other fixed-sized field. 
We show that well-known matroid-theoretic criteria on binary linear codes give conditions on the lattice of cyclic flats, which govern the locality properties of the associated storage codes. As a consequence, we get stronger structural constraints on binary LRCs  compared to those previously known for general LRCs. Moreover, a collection of criteria for detecting uniform minors from the lattice of cyclic flats of a given matroid is given, which is  interesting in its own right.

\section{Preliminaries on LRCs and Matroids}

To study LRCs in more detail, we consider punctured codes $C|Y$, where $Y\subseteq E$ is a set of coordinates of the code $C$. For a fixed code $C$, we denote by $d_Y$ the minimum Hamming distance of the punctured code $C|Y$.
%
%
%
As is common practice, we say that $C$ is an $(n,k,d)$-code if it has length $n$, dimension $k$ and minimum Hamming distance $d$. 
%
A linear $(n,k,d)$-code $C$ over a field is a \emph{non-degenerate storage code} if $d \geq 2$ and there is no zero column in a generator matrix of $C$.

\begin{definition}
Here, a \emph{linear $(n,k,d,r,\delta)$-LRC} over a finite field $\F$ is a non-degenerate linear $(n,k,d)$-code $C$ over $\F^E$ such that any coordinate $x \in E$ of $C$ has \emph{locality} $(r,\delta)$, meaning that there is a subset $R$ of $E$, called \emph{repair set} of $x$, such that $x \in R$, $|R|\leq r+\delta-1$ and $d_R\geq \delta$.
\end{definition}

The parameters $(n,k,d,r,\delta)$ can immediately be defined and studied for matroids in general, as in~\cite{tamo13,westerback15}. 

\subsection{Matroid fundamentals} 
\label{dualization}

Matroids were first introduced by Whitney in 1935, to capture and generalise the notion of linear dependence in purely combinatorial terms. 
Indeed, the combinatorial setting is general enough to also capture many other notions of dependence occurring in mathematics, such as cycles or incidences in a graph, non-transversality of algebraic varieties, or algebraic dependence of field extensions. Of special interest for linear LRCs is the connection between linear algebra and matroids. 

Matroids have many equivalent definitions in the literature. Here, we choose to present matroids via their rank functions. Much of the contents in this section can be found in more detail in \cite{freij-hollanti17}.

\begin{definition}[Matroid]\label{def:matroid_rank}
A \emph{(finite) matroid} $M=(\rho,E)$ is a finite set $E$ together with a \emph{rank function} $\rho:2^E \rightarrow \mathbb{Z}$ such that for all subsets $X,Y \subseteq E$
$$
\begin{array}{rl} 
(R.1) & 0 \leq \rho(X) \leq |X|,\\
(R.2) & X \subseteq Y \quad \Rightarrow \quad \rho(X) \leq \rho(Y),\\
(R.3) & \rho(X) + \rho(Y) \geq \rho(X \cup Y) + \rho(X \cap Y). 
\end{array}
$$ 
\end{definition}

A subset $X \subseteq E$ is called \emph{independent} if $\rho(X) = |X|$. If $X$ is independent and $\rho(X) = \rho(E)$, then $X$ is called a \emph{basis}. Strongly related to the rank function is the \emph{nullity function} $\eta:2^E \rightarrow \mathbb{Z}$, defined by $\eta(X) = |X| - \rho(X)$ for $X \subseteq E$.

Any matrix $G$ over a field $\F$ generates a matroid $M_G=(\rho,E)$, where $E$ is the set of columns of $G$, and $\rho(X)$ is the rank of $G(X)$ over $\F$, where $G(X)$ denotes the submatrix of $G$ formed by the columns indexed by $X$. As elementary row operations preserve the row space of $G(X)$ for all $X\subseteq E$, it follows that row-equivalent matrices generate the same matroid.  

Thus, there is a straightforward connection between linear codes and matroids. Let $C$ be a linear code over a field $\F$. 
Then any two different generator matrices of $C$ will have the same row space by definition, so they will generate the same matroid. Therefore, without any inconsistency, we can denote the matroid associated to these generator matrices by $M_C = (\rho_C,E)$. The rank function $\rho_C$ can be defined directly from the code without referring to a generator matrix, via $\rho_C(X) = \mathrm{dim}(C|X)$ for $X \subseteq E$.
\begin{example} \label{ex:code_matroid}
Let $C$ be the linear code generated by the following matrix $G$ over $\F_{2}$:

\[
\begin{small}
G=
\begin{tabular}{ |c|c|c|c|c|c| }
\multicolumn{1}{c}{1}&
\multicolumn{1}{c}{2}&
\multicolumn{1}{c}{3}&
\multicolumn{1}{c}{4}&
\multicolumn{1}{c}{5}&
\multicolumn{1}{c}{6}\\
\hline
1&0&1&0&1&1\\
\hline
0&1&1&0&1&1\\
\hline
0&0&0&1&1&1\\
\hline
\end{tabular}
\end{small}
\]

Then, for the matroid $M_{C}= (\rho_{C}, \{1,2,3,4,5,6\})$,
\[
\rho_{C}(\emptyset) = 0, \;  \rho_{C}(\{1,2,3\}) = \rho_{C}(\{ 3,4,5 \})= 2, \; \rho_{C}(\{1,2,3,4,5,6\})=3.
\]
\end{example}

Two matroids $M_1 = (\rho_1, E_1)$ and $M_2 = (\rho_2,E_2)$ are \emph{isomorphic} if there exists a bijection $\psi: E_1 \rightarrow E_2$ such that $\rho_2(\psi(X)) = \rho_1(X)$ for all subsets $X \subseteq E_1$.

\begin{definition} A matroid that is isomorphic to $M_G$ for some matrix $G$ over $\F$ is said to be \emph{representable} over $\F$. We also say that such a matroid is $\F$-representable. 
A \emph{binary} matroid is a matroid that is $\F_2$-representable.\end{definition}

\begin{definition}
The \emph{uniform matroid} $U_n^k=(\rho, [n])$ is a matroid with a ground set $[n]=\{1,2,\ldots , n\}$ and a rank function $\rho(X)=\min\{|X|, k\}$ for $X\subseteq [n]$.  
\end{definition}


The following straightforward observation gives a characterisation of maximum distance separable (MDS) codes and also shows that uniform matroids constitute a subclass of representable matroids.

\begin{proposition}
A linear code $C$ is an $(n,k,n-k+1)$-MDS code if and only if $M_C$ is the uniform matroid $U_n^k$.
\end{proposition}

There are several elementary operations that are useful for explicit constructions of matroids, as well as for analysing their structure. The operations that we will need for this paper are dualisation, contraction,  and deletion.

\begin{definition}
Let $M = (\rho,E)$ be a matroid and $X, Y \subseteq E$, and denote by $\bar{X}=E-X$ for any $X\subseteq E$. Then
\begin{enumerate}[(i)]
\item The \emph{restriction} of $M$ to $Y$ is the matroid $M|Y = (\rho_{|Y}, Y)$, where $\rho_{|Y}(A) = \rho(A)$ for $A \subseteq Y$.
\item The \emph{contraction} of $M$ by $X$ is the matroid $M/X = (\rho_{/X}, \bar{X})$, where $\rho_{/X}(A) = \rho(A \cup X) - \rho(X)$ for $A \subseteq \bar{X}$.
\item A \emph{minor} of $M$ is the matroid $M|Y/X = (\rho_{|Y/X}, Y-X)$ obtained from $M$ by restriction to $Y$ and contraction by $X$. Observe that this does not depend on the order in which the restriction and contraction are performed.
\item The \emph{dual} of $M$ is the matroid $M^* = (\rho^*, E)$, where $\rho^*(A) = |A| + \rho(\bar{A}) - \rho(E)$ for $A \subseteq E$.
\end{enumerate}
\end{definition}

The restriction operation to $Y$ is also referred to as \emph{deletion} of the set $E-Y$. 

Let $M=M_C$ be a representable matroid. Then, restriction, contraction, and dualisation of $M$ correspond to puncturing, shortening, and orthogonal complement of the code $C$, respectively. 

\begin{example}\label{ex:LCF}
Let $C$ be the code generated by the matrix $G$ in Example \ref{ex:code_matroid}. Then for the matroid $M_{C} =(\rho = \rho_C,[6])$, $X = \{ 1 \}$, and $Y = \{ 2,3,4,5 \}$, 
\[
\rho_{/X}(\{ 2,3,4 \}) = 2, \; \rho_{|Y}(\{2,3,4 \}) = 3, \hbox{ and } \rho^{*}( \{ 2,3,4\} ) =2. 
\]

\end{example}

The minors of uniform matroids are very easily described: 
\begin{lemma}\label{lm:subunif}
Let $U_n^k=(\rho, [n])$ be a uniform matroid, and let $X\subseteq Y\subseteq E$. Then the minor $U_n^k|Y/X$ is isomorphic to $U_{n'}^{k'}$, where $k'=\max\{0,k-|X|\}$ and $n'=|Y|-|X|$. In particular, $M$ is a minor of $U_n^k$ if and only if $M\cong U_{n'}^{k'}$, for some $0\leq k'\leq k$ and $0\leq n'-k'\leq n-k$.
\end{lemma}

In general there is no simple criterion to determine if a matroid is representable.  However, there is a simple criterion for when a matroid is binary. 

\begin{theorem}[\cite{tutte58}]\label{tutte}
Let $\rmatroid$ be a matroid. The following two conditions are equivalent.
\begin{enumerate}
\item $M$ is linearly representable over $\F_2$.
\item There are no sets $X\subseteq Y\subseteq E$ such that $M|Y/X$ is isomorphic to the uniform matroid $U_4^2$.
\end{enumerate}
\end{theorem}

In essence, this means
that the only obstruction that needs to be overcome in order to be representable over the binary
alphabet, is that no more than three nonzero points can fit in the same plane. Clearly, if $M$ is representable over $\F$, then so are all its minors. The following result, Theorem \ref{thm:rota}, is a far-going extension of Theorem~\ref{tutte}, that was first conjectured by Gian-Carlo Rota in 1970.
A proof of this conjecture was announced by Geelen, Gerards, and Whittle in 2014, but the details of the proof still remain to written up~\cite{whittle14}. 

\begin{theorem}[\cite{whittle14}]\label{thm:rota}
For any finite field $\F$, there is a finite set $L(\F)$ of matroids such that any matroid $M$ is representable if and only if it contains no element from $L(\F)$ as a minor.\end{theorem}

Since the 1970's,
it has been known that a matroid is representable over $\F_3$ if and only if it avoids the uniform
matroids $U_5^2$, $U_5^3$, the Fano plane $\P^2(\F_2)$, and its dual $\P^2(\F_2)^*$ as minors. The list $L(\F_4)$ has seven elements, and was given explicitly in 2000. For larger fields, the explicit list is not known, and there is little hope to even find useful bounds on its size.

By the Critical Theorem \cite{crapo70}, the matroid $M_C$ determines the supports of a linear code $C$. Consequently, since binary codes are determined uniquely by the support of the codewords, binary matroids are in one-to-one correspondence with binary codes. This is in sharp contrast to linear codes over larger fields, where many interesting properties are not determined by the associated matroid. 
An important example of such a property is the covering radius \cite{britz05}. 

\subsection{Fundamentals on cyclic flats}

The main tool from matroid theory in this paper are the cyclic flats. We will define them using the closure and cyclic operator.   

Let $M = (\rho,E)$ be a matroid. The \emph{closure} operator $\mathrm{cl}:2^E \rightarrow 2^E$ and \emph{cyclic} operator $\mathrm{cyc}: 2^E \rightarrow 2^E$ are defined by
$$
\begin{array}{cl}
(i) & \mathrm{cl}(X) = X \cup \{e \in E -X : \rho(X \cup e) = \rho(X)\},\\
(ii) & \mathrm{cyc}(X) = \{e \in X : \rho(X - e) = \rho(X)\}.
\end{array}
$$
A subset $X \subseteq E$ is a \emph{flat} if $\mathrm{cl}(X) = X$ and a \emph{cyclic set} if $\mathrm{cyc}(X) = X$. Therefore, $X$ is a \emph{cyclic flat} if 
$$
\rho(X \cup y) > \rho(X) \quad \hbox{and} \quad \rho(X - x) = \rho(X)
$$
for all $y \in E-X$ and $x \in X$. The collection of flats, cyclic sets, and cyclic flats of $M$ are denoted by $\mathcal{F}(M)$, $\mathcal{U}(M)$, and $\mathcal{Z}(M)$, respectively. 

It is easy to verify, as in~\cite{bonin08}, that the closure operator induces flatness and preserves cyclicity, and that the cyclic operator induces cyclicity and preserves flatness. Thus we can write \[\cl: \left\{\begin{split} 2^E & \to\FF(M)\\ \UU(M)&\to\ZZ(M),\end{split}\right.\quad\textrm{ and }\quad\cyc: \left\{\begin{split} 2^E & \to\UU(M)\\ \FF(M)&\to\ZZ(M).\end{split}\right.\] In particular, for any set $X\subseteq E$, we have $\cyc(\cl(X)) \in\ZZ(M)$ and $\cl(\cyc(X)) \in\ZZ(M)$.
Some more fundamental properties of flats, cyclic sets, and cyclic flats are given
 in \cite{bonin08}. 

The cyclic flats of a linear code $C$ of $\F^E$ can be described as sets $X \subseteq E$ such that 
$$
C|(X \cup y) \supsetneq C|X \quad \hbox{and} \quad C|(X - x) = C|X
$$ 
for all $y \in E -X$ and $x \in X$. Thus, we have the following immediate proposition.

\begin{proposition}\label{prop:uniform}
Let $M = (\rho,E)$ be a matroid. The following are equivalent:
\begin{enumerate}[$(i)$]
\item $M$ is the uniform matroid $U_n^k$
\item $\ZZ = \ZZ(M)$ is the two element lattice with bottom element $0_{\ZZ}=\emptyset$, top element $1_{\ZZ} = E$ and $\rho(1_{\ZZ}) = k$
\end{enumerate}

\end{proposition} 

Non-degeneracy of a truncated code can be observed immediately from the lattice of cyclic flats of the associated matroid, as follows.

\begin{proposition}
Let $C$ be a linear code over $\F^E$ and $X \subseteq E$. Then $C|X$ is non-degenerate if and only if $1_{\mathcal{Z}(M_C|X)} = X$ and $0_{\mathcal{Z}(M_C|X)} = \emptyset$. 
\end{proposition}

The minimum distances and the ranks of punctured codes can be computed from the lattice of cyclic flats via the following theorem. In~\cite{westerback15}, this was used to construct matroids and linear LRCs with prescribed parameters $(n,k,d,r,\delta)$.

\begin{theorem}[\cite{westerback15}]
Let $C$ be a linear code over $\F^E$ and $X \subseteq E$. Then, if $C|X$ is non-degenerate, it has dimension $k_X=\rho(1_{\mathcal{Z}(M_C|X)})$ and minimum distance $d_X=\eta(X) + 1 -  \max \{\eta(Y) : Y \in \mathcal{Z}(M_C|X) \hbox{ and } Y \neq X \}$.
\end{theorem}

%

\begin{example} 
Let $M_C= (\rho_{C}, E=[6])$ be the matroid associated to the linear code $C$ generated by the matrix $G$ given in Example \ref{ex:code_matroid}. The lattice of cyclic flats $(\ZZ, \subseteq )$ of $M_{C}$ is given in Fig. \ref{fig:LCF_Ex}, where the cyclic flat is given inside the node and its rank is labelled outside the node on the right.
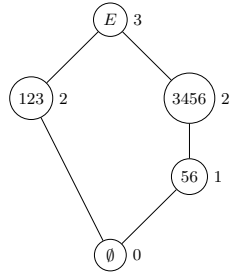
\begin{figure}[hb!]
\centering
\resizebox{0.25\textwidth}{!}{%
\begin{tikzpicture}
	\node[shape=circle,draw=black, label=right:0] (00) at (0,0) {\small$\emptyset$};  
    \node[shape=circle,draw=black, label=right:1] (12) at (1.5,1.5) {\small\(56\)}; 
    \node[shape=circle,draw=black, label=right:2] (21) at (-1.5,3) {\small\(123\)}; 
    \node[shape=circle,draw=black, label=right:2] (22) at (1.5,3) {\small \(3456\)}; 
    \node[shape=circle,draw=black, label=right:3] (31) at (0,4.5) {\small \(E \)};
     \path [-] (00) edge (12);
    \path [-] (00) edge (21);
    \path [-] (12) edge (22);
    \path [-] (21) edge (31);
    \path [-] (22) edge (31);
\end{tikzpicture}
}
\caption{Lattice of cyclic flats of $M_{C}$.}
\label{fig:LCF_Ex}
\end{figure}

From the lattice of cyclic flats given above, we can conclude that $C$ is a $(6, 3, 2)$-code.

\end{example}

\section{Sufficient Conditions for Uniformity}
Our main goal is to study criteria for when $M|Y/X$ is uniform, for $X\subseteq Y\subseteq E(M)$. As a preparation, observe that for a uniform matroid $M\cong U_n^k$, the only cyclic flats are $\emptyset$ and $E(M)$, with $|E(M)|=n$ and $\rho(E)=k$. 

Our first interest is in  the case when $X$ and $Y$ are themselves cyclic flats. Then we have a straightforward characterisation of cyclic flats in $M|Y/X$, via the following two lemmas:

\begin{lemma}\label{flats}
Let $M$ be a matroid, and let $X\subseteq Y\subseteq E(M)$ be two sets with $Y\in\FF(M)$. Then $\FF(M|Y/X)=\{F\subseteq Y-X , F\cup X\in\FF(M)\}$.
\end{lemma}

\begin{proof}
A set $S$ is flat in $M|Y/X$ precisely if $\rho(S\cup X\cup i)>\rho(S\cup X)$ for all $i\in (Y-X)- S$. Since $Y$ is flat, the inequality $\rho(S\cup X\cup i)>\rho(S\cup X)$ will hold for all $i\in \bar{Y}$ regardless of $S$. Thus, $S$ is flat in $M|Y/X$ if and only if $S\cup X$ is flat in $M$.
\end{proof}

\begin{lemma}\label{cyclic}
Let $M$ be a matroid, and let $X\subseteq Y\subseteq E(M)$ be two sets with $X\in\UU(M)$. Then $\UU(M|Y/X)=\{U\subseteq Y-X , U\cup X\in\UU(M)\}$.
\end{lemma}
This is the dual statement, and thus an immediate consequence, of Lemma~\ref{flats}. We write out the proof explicitly only for illustration.
\begin{proof}
A set $S$ is cyclic in $M|Y/X$ precisely if $\rho((S\cup X)- i)=\rho(S\cup X)$ for all $i\in S$. For $i\in X$, this will hold regardless of $S$, since $X$ is cyclic. Thus, $S$ is cyclic in $M|Y/X$ if and only if $S\cup X$ is cyclic in $M$.
\end{proof}

The previous lemmas give the following immediate corollary:
\begin{corollary}\label{corr:interval}
Let $M=(E, \rho)$ be a matroid, and let $X\subseteq Y\subseteq E(M)$ be two sets with $X\in\UU(M)$ and $Y\in \FF(M)$. Then $\ZZ(M|Y/X)=\{Z\subseteq Y-X , Z\cup X\in\ZZ(M)\}$, with the rank function $\rho_{|Y/X}(Z)=\rho(Z\cup X)-\rho(X)$.
\end{corollary}

As a consequence, we get a sufficient condition for uniformity of minors, that only depends on the Hasse diagram of $\ZZ(M)$. Recall that $v$ is said to \emph{cover} $u$ in a poset $P$ if $u<v$ and there is no $w$ with $u<w<v$. If this is the case, then we write $u\lessdot_P v$. This is equivalent to $(u,v)$ being an upwards directed edge in the Hasse diagram of $P$.

\begin{theorem}\label{thm:edge}
Let $X$ and $Y$ be two cyclic flats in $M$ with $X\lessdot_{\ZZ(M)} Y$. Let $n=|Y|-|X|$ and $k=\rho(Y)-\rho(X)$. Then $M|Y/X\cong U_n^k$.
\end{theorem}

\begin{proof}
By Corrolary~\ref{corr:interval}, we have $\ZZ(M|Y/X)=\{\emptyset,Y-X\}$ as there are no cyclic flats with $X\subset Z\subset Y$. Again by Corrolary~\ref{corr:interval}, we have $\rho(Y-X)=\rho(Y)-\rho(X)=k$. Thus, by Proposition~\ref{prop:uniform}, $M|Y/X$ is the uniform matroid $U_n^k$.
\end{proof}

\begin{corollary}\label{corr:hasse}
Let $M$ be a matroid that contains no $U_n^k$ minors. Then, for every edge $X\lessdot_{\ZZ(M)} Y$ in the Hasse diagram of $\ZZ(M)$, we have $\rho(Y)-\rho(X)<k$ or $\eta(Y)-\eta(X)<n-k$.
\end{corollary}
\begin{proof}
Assume for a contradiction that $X\lessdot_{\ZZ(M)} Y$ has $\rho(Y)-\rho(X)=k'\geq k$ or $\eta(Y)-\eta(X)=n'-k'\geq n-k$. Then by Theorem~\ref{thm:edge}, $M|Y/X\cong U_{n'}^{k'}$, and so contains $U_n^k$ as a minor by Lemma~\ref{lm:subunif}.
\end{proof}

Now, we are going to need formulas for how to compute the lattice operators in $\ZZ(M|Y/X)$ in terms of the corresponding operators in $\ZZ(M)$. These can be derived from corresponding formulas for the closure and cyclic operator. To derive these, we will need to generalise Corollary~\ref{corr:interval} to the setting where the restriction and contraction are not necessarily performed at cyclic flats.

\begin{theorem}\label{thm:cf_formulas}
For $X \subseteq Y \subseteq E$, we have
\begin{enumerate}
\item $\ZZ(M|Y) = \{ \cyc(Z \cap Y) : Z \in \ZZ(M) \}$
\item $\ZZ(M/X) = \{ \cl(X \cup Z ) - X : Z \in \ZZ(M) \}$
\item $\ZZ(M|Y/X) = \left\lbrace \cl\Big( X \cup \cyc \big( Z \cap Y\big) \Big) \cap \big(Y-X \big) : Z \in \ZZ(M) \right\rbrace \\
= \left\lbrace \cyc\Big( \cl \big(X \cup Z \big) \cap Y \Big) - X : Z \in \ZZ(M) \right\rbrace$
\end{enumerate}
\end{theorem}

\begin{proof}
\begin{enumerate}
\item First, observe that the cyclic operator in $M|Y$ is the same as that in $M$, and that the flats in $M|Y$ are $\{ F \cap Y : F \in \FF(M) \}$. Thus we have \[\ZZ(M|Y)=\{\cyc(F\cap Y) : F\in\FF(M)\}\supseteq \{\cyc(Z\cap Y) : Z\in\ZZ(M)\}.\]  
On the other hand, let $A \in \ZZ(M|Y)$, so $\cyc(A) = A$ and $\cl(A) \cap Y = A$. But the closure operator preserves cyclicity, so $\cl(A)\in\ZZ(M)$. We then observe that \[A = \cyc(\cl(A) \cap Y)\in \{\cyc(Z\cap Y) : Z\in\ZZ(M)\}.\] This proves the reverse inclusion \[\ZZ(M|Y)=\{\cyc(F\cap Y) : F\in\FF(M)\}\subseteq \{\cyc(Z\cap Y) : Z\in\ZZ(M)\}.\]  

\item This is the dual statement of 1., and so follows immediately by applying 1. to the matroid $M^*|\bar{X}/\bar{Y}$. 
%

\item We first apply 2. and then 1. to the restricted matroid $M|Y$, and get
\[
\ZZ(M|Y/X) = \{ \cl_{|Y}(X \cup \cyc(Z \cap Y) ) - X : Z \in \ZZ(M) \}.
\]
But if $T \subseteq Y$, then $\cl_{|Y}(T) = \cl(T) \cap Y$. Then,
\[
\ZZ(M|Y/X) = \{ \cl(X \cup \cyc(Z \cap Y)) \cap (Y - X) : Z \in \ZZ(M) \}
\]

For the second equality of 3. we need to study the operator $\cyc_{/X}$. Suppose $T \subseteq E-X$. Using duality and the formula for $cl_{| \bar{X}}$, we find that 
\[
\cyc_{/X}(T) = \cyc(X \cup T ) - X.
\]


Now we are ready to prove the last equality. Applying first 1. and then 2. to the contracted matroid $M/X$, we get
\[
\ZZ(M|Y/X) = \{ \cyc_{/X} ( ( \cl(X \cup Z ) - X ) \cap Y ) : Z \in \ZZ(M) \}
\]
Applying the formula for $\cyc_{/X}$, we obtain
\begin{align*}
\ZZ(M|Y/X) & = \{ \cyc(( \cl(X \cup Z ) \cap Y-X ) \cup X ) - X : Z \in \ZZ(M) \}\\ & = \{ \cyc ( \cl( X\cup Z) \cap Y ) - X : Z \in \ZZ(M) \},
\end{align*}
where the last equality follows as $X \subseteq cl(X \cup Z )$. This concludes the proof.
\end{enumerate}
\end{proof}

\section{Criteria for Uniformity via Cyclic Flats}




We can use Theorem \ref{thm:cf_formulas} to find conditions for a minor to be isomorphic to a uniform matroid. The idea is to detect when $\ZZ(M|Y/X)$ as calculated in Theorem \ref{thm:cf_formulas} is precisely $\{\emptyset, Y-X\}$. Using this, we will be able to find some conditions on the sets $X$ and $Y$, as well as on the matroid itself.

\subsection{Minors Given by Restriction or Contraction Only}

We will begin by considering a simpler case when the minor is the result of a restriction only, \emph{i.e.,} when the minor is given by $M|Y$. So, let $M=(E, \rho)$ be a matroid and $Y$ an arbitrary subset of $E$. We can use Corollary \ref{corr:interval} to restrict the amount of information we need to consider. Indeed, by properties of minors, we have
\begin{equation}
\label{minor_1}
 M|Y = M|\cl(Y) \setminus (\cl(Y)-Y).
\end{equation} 
Then, Theorem \ref{thm:cf_formulas} states that the cyclic flats of $M|Y$ depend only on the cyclic flats of $M|\cl(Y)$. Furthermore, according to Corollary \ref{corr:interval} the cyclic flats of $M|\cl(Y)$ are exactly the cyclic flats of $M$ contained in $\cl(Y)$. Hence, we can restrict the study to the case when $M=(E, \rho)$ is a matroid and $Y$ is a subset of full rank. Define $k:=\rho(Y)$ and $n:=|Y|$. With this setup, we obtain the following theorem.

\begin{theorem}
\label{thm_restriction_case}
Let $M=(E,\rho)$ be a matroid and $Y$ a subset of full rank. $M|Y$ is isomorphic to the uniform matroid $U_n^k$ if and only if either $Y$ is a basis of $M$ or the following two conditions are satisfied:
\begin{enumerate}
\item $Y$ is a cyclic set of $M$.
\item For all $Z \in \ZZ(M)$ with $\rho(Z)<k$, we have that $Z \cap Y$ is independent in $M$.  
\end{enumerate}
\end{theorem}

Before stating the proof, we will need one useful lemma about the properties of the closure and cyclic operator. 

\begin{lemma}
\label{lemma:cl_cyc}
Let $M=(E,\rho)$ be a matroid and $Y \subseteq E$. Then
\begin{enumerate}
\item $\cl(\cyc(Y)) \cap Y = \cyc(Y)$.
\item $\cyc(\cl(Y)) \cup Y = \cl(Y)$.
\end{enumerate}
\end{lemma}

The proof of Lemma~\ref{lemma:cl_cyc} is straightfrorward from the definition of the operators together with the submodularity of the rank function. Details of the proof can be found in~\cite{bonin08}. We now present the proof of Theorem \ref{thm_restriction_case}. 

\begin{proof}
We know that $M|Y \cong U_n^k$ if and only if $\ZZ(M|Y) = \{ \emptyset, Y \}$. On the other hand, we know by Theorem \ref{thm:cf_formulas}, that $\ZZ(M|Y) = \{ \cyc(Z \cap Y ) : Z \in \ZZ(M) \}$. Then we have
\[
M|Y \cong U_n^k \text{ if and only if } \cyc(Z \cap Y) \in \{ \emptyset, Y \} \text{ for all } Z \in \ZZ(M). 
\]

Now consider the cyclic flat $Z_{Y} := \cl(\cyc(Y))$. Using Lemma \ref{lemma:cl_cyc}, we have $\cyc(Z_{Y} \cap Y) = \cyc(Y)$. Two cases can occur. If $\cyc(Y) = \emptyset$ then $Y$ was a basis of $M$ and we end up with a minor isomorphic to $U_k^k$. If not, then $\cyc(Y)$ must be equal to $Y$. So, we have that $Y \in \ZZ(M|Y)$ if and only if $Y$ is a cyclic set and we obtain the first condition. Since $Y$ already has full rank, there is only one cyclic flat that contains $Y$, namely $\cl(\cyc(Y)) = E$. Therefore, for every other cyclic flat $Z$, \emph{i.e.,} for all $Z$ with $\rho(Z)<k$, we have $\cyc(Z \cap Y) \subseteq Z \cap Y \neq Y$. But, by Theorem \ref{thm:cf_formulas}, $\cyc(Z \cap Y ) $ is a cyclic flat of $M|Y$. Thus, for all $Z \in \ZZ(M)$ with $\rho(Z) <k$, we have $\cyc(Z\cap Y) = \emptyset$, or equivalently, $Z \cap Y$ is independent. Notice that, combined with the first condition, this implies immediately that $\ZZ(M|Y) = \{ \emptyset, Y \}$. This concludes the proof. 
\end{proof}

\begin{corollary}
Under the above assumptions, if $M|Y$ is isomorphic to $U_n^k$ then the ground set $E$ must be a cyclic flat, \emph{i.e.,} $E \in \ZZ(M)$. 
\end{corollary}

\begin{example}

We will look at the matroid $M_{C}$ arising from the binary matrix in Example \ref{ex:code_matroid}. Since it is a binary matroid, we cannot find a minor isomorphic to $U_4^2$. Using the lattice of cyclic flats of Fig. \ref{fig:LCF_Ex}, we can find a minor isomorphic to $U_3^2$ if we look at, \emph{e.g.},  $M_{C}|\{1,2,3\}$. We can also find a minor isomorphic to $U_4^3$. To this end, we look at $M_{C}|\{1,2,4,5\}$. The set $\{1,2,4,5\}$ is indeed a cyclic set because $\cyc(\{1,2,4,5 \} ) = \{1,2,4,5\}$. Furthermore, there is no cyclic flat properly contained in $\{1,2,4,5\}$, and every intersection of a cyclic flat different from $E$ with the set $\{1,2,4,5 \}$ gives us an independent set. Thus, the conditions from the previous theorem are met and we get a minor isomorphic to $U_4^3$. 

\end{example}

Now we can do the same for $M/X$ and use duality to get back to the restriction case.  By minor properties, we have
\begin{equation}
\label{minor_2}
 M/X = M/\cyc(X) / (X-\cyc(X)).
\end{equation}
 
Then, Corollary \ref{corr:interval} states that the cyclic flats of $M/\cyc(X)$ are the cyclic flats of $M$ that contain $\cyc(X)$. Thus, we will consider a matroid $M=(E,\rho)$ and $X$ an independent subset of $E$. Define $k:=\rho(E)-\rho(X)$ and $n:=|E-X|$. Then, we have the following dual statement of Theorem \ref{thm_restriction_case}. 

\begin{theorem}
Let $M=(E,\rho)$ be a matroid and $X$ an independent subset of $E$. $M/X$ is isomorphic to the uniform matroid $U_n^k$ if and only if either $X$ is a basis of $M$ or the following two conditions are satisfied:
\begin{enumerate}
\item $X$ is a flat of $M$.
\item For all $Z \in \ZZ(M)$ with $\rho(Z)>0$, we have $\cl(X \cup Z ) = E$.
\end{enumerate}
\end{theorem}

\begin{corollary}
Under the above assumptions, if $M/X$ is isomorphic to $U_n^k$ then the empty set must be a cyclic flat, \emph{i.e.,} $\emptyset \in \ZZ(M)$.
\end{corollary}

\subsection{Minors Given by Both Restriction and Contraction}

This part combines the two previous situations into a more general statement. We will see that, when we allow both a restriction and a contraction to occur, we lose some conditions on the matroid that are then replaced by conditions on the sets used in the minor.


Let $M=(E,\rho)$ be a matroid and $X \subset Y \subseteq E$ two sets. Combining minor properties (\ref{minor_1}) and (\ref{minor_2}) and Corollary \ref{corr:interval}, it is sufficient to only consider the cyclic flats  between $\cyc(X)$ and $\cl(Y)$. In addition, we want to avoid some known cases, namely when $Y$ is a basis (we will obtain $U_k^k$) and when $X$ has full rank (we will obtain $U_0^0$). Define $k:=\rho(E)-\rho(X)$ and $n:=|Y-X|$. We get the following theorem.  

\begin{theorem}
Let $M=(E,\rho)$ be a matroid and $X \subset Y \subseteq E$ two sets such that $Y$ is a dependent full-rank set and $X$ is an independent set with $\rho(X)<\rho(E)$. The minor $M|Y/X$ is isomorphic to a uniform matroid $U_n^k$ if and only if
\begin{enumerate}
\item $\cl(X) \cap Y = X$,
\item $Y-X \subseteq \cyc(Y)$,
\item for all $Z \in \ZZ(M)$ either $Z \cap Y$ is independent or\\ $\cl(X \cup \cyc(Z \cap Y)) = E$.
\end{enumerate}
\end{theorem}

\begin{proof}
Using Theorem \ref{thm:cf_formulas}, we have that $M|Y/X \cong U_n^k$ if and only if 
\[
\cl(X \cup \cyc(Z \cap Y) ) \cap (Y-X) \in \{ \emptyset, Y-X \} \text{ for all } Z \in \ZZ(M).
\]
In particular, it holds for $Z = 0_{\ZZ}$. Let $Z'_{0} := \cl(X \cup \cyc(0_{\ZZ} \cap Y))$. Using the properties of the closure, we have
\[
\cl(X) \subseteq Z'_{0} \subseteq \cl(X \cup 0_{\ZZ} ) = \cl(X).
\]
Thus, we have a chain of equalities and, in particular, $\rho(Z'_{0}) < \rho(E)$. This means that $Z'_{0} = \emptyset$ in order to have $\emptyset \in \ZZ(M|Y/X)$. But, since $Z'_{0} = \cl(X)$ then $Z'_{0} = \emptyset$ is equivalent to  $\cl(X) \cap Y = X$ and Condition~1 is proved.

Now, consider the cyclic flat $Z_{Y} := \cl(\cyc(Y))$. First, using again Lemma \ref{lemma:cl_cyc}, we have $\cyc(Z_{Y} \cap Y ) = \cyc(Y)$. Since $Y$ is a dependent subset, $\cyc(Y) \neq \emptyset$ and thus $X \subsetneq X \cup \cyc(Y)$. Then, the closure cannot be contained in $X$ and we must have
\[
\cl(X \cup \cyc(Y) ) \cap (Y-X) = Y-X.
\]

Define $Z'_{Y} := \cl(X \cup \cyc(Y) )$. The above equality means that $Y-X \subseteq Z'_{Y}$. On the other hand, we have that $X \subseteq Z'_{Y}$. Then, $Y \subseteq Z'_{Y}$ and $Z'_{Y} = E$. In particular, we must have $Y-X \subseteq \cyc(Y)$. Indeed, assume by contradiction that there exists $a \in Y-X$ and $a \notin \cyc(Y)$. Then, by definition of the cyclic operator, $\rho(Y-a)<\rho(Y)$. But since $a \notin \cyc(Y)$ and $a \notin X$ then $X \cup \cyc(Y) \subseteq Y-a$. This implies that $\rho(X \cup \cyc(Y)) \leq \rho(Y-a) < \rho(Y)$ which is a contradiction. The condition $Y-X \subseteq \cyc(Y)$ is also sufficient to guarantee that $Y-X \in \ZZ(M|Y/X)$. This proves Condition 2.  

Finally, for every other cyclic flat of $M$, $\cl(X \cup \cyc(Z \cap Y) ) \cap (Y-X) = \emptyset$ if and only if $\cyc(Z \cap Y) \subseteq X$. But since $X$ is independent, this is equivalent to $Z \cap Y$ being independent. On the other hand, $\cl(X \cup \cyc(Z \cap Y) ) \cap (Y-X) = Y-X$ if and only if $\cl(X \cup \cyc(Z \cap Y) ) = E$. This concludes the proof. 
\end{proof}

\section{Stuctural Properties of Binary LRCs}
From a practical viewpoint, storage systems over alphabets of bounded size are of special interest. The field size is important both because it governs the complexity of the computations involved in repair and retrieval, and because it restricts the size of the data items stored. We are therefore interested in understanding the matroidal structure of LRCs that are linearly representable over the finite field $\F_q$, where $q$ is small.

Assuming the MDS conjecture~\cite{Segre55}, a matroid $M$ that is linearly representable over $\F_q$ must avoid $U_{q+2}^k$ as a minor, for $k=2$, $4\leq k\leq q-2$, and $k=q$. If $q$ is odd, $M$ must also avoid $U_{q+2}^3$ and $U_{q+2}^{q-1}$ minors. The MDS conjecture is widely believed to be true, and is proven when $q$ is prime~\cite{BallMDS}. 

By Corollary~\ref{corr:hasse}, matroids avoiding $U_n^k$ minors have a rather special structure in their lattice of cyclic flats. In particular, it tells us that whenever $X\lessdot_\ZZ Y$, we cannot simultaneously have $\rho(Y)-\rho(X)>k$ and $\eta(Y)-\eta(X)>n-k$. This observation can be exploited to bound locality parameters of an LRC in terms of the field size.

Of special interest are binary storage codes that are linear over $\F_2$. It is known that a matroid is representable over $\F_2$ if and only if it avoids $U_4^2$ as a minor. In this case,  Corollary~\ref{corr:hasse} tells us that we cannot simultaneously have $\rho(Y)-\rho(X)>1$ and $\eta(Y)-\eta(X)>1$. 
On the other hand, we know by Theorem 3.2 in \cite{bonin08} or by direct calculation, that we always have $\rho(Y)-\rho(X)\geq 1$ and $\eta(Y)-\eta(X)\geq 1$. Thus, if $M$ is representable over $\F_2$, then every edge $X\lessdot_\ZZ Y$ in the Hasse diagram of $\ZZ(M)$ satisfies exactly one of the following:

\begin{enumerate}[(i)]
\item $\rho(Y)-\rho(X)=l>1$. We call such an edge a \emph{rank edge}, and label it $\rho=l$. Such an edge corresponds to a $U_{l+1}^l$ minor in $M$.
\item $\eta(Y)-\eta(X)=l>1$. We call such an edge a \emph{nullity edge}, and label it $\eta=l$. Such an edge corresponds to a $U_{l+1}^1$ minor in $M$.
\item $\rho(Y)-\rho(X)=1$ and $\eta(Y)-\eta(X)=1$. We call such an edge an \emph{elementary edge}. Such an edge corresponds to a $U_2^1$ minor in $M$.
\end{enumerate}
As an example, the matroid from Examples~\ref{ex:code_matroid} and~\ref{ex:LCF} gets an edge labelling as illustrated in Figure~\ref{fig:LCF_ExR}:

\begin{figure}

\centering

\resizebox{0.25\textwidth}{!}{%
\begin{tikzpicture}

	\node[shape=circle,draw=black] (00) at (0,0) {\small$\emptyset$};  

    \node[shape=circle,draw=black] (12) at (1.5,1.5) {\small\(56\)}; 

    \node[shape=circle,draw=black] (21) at (-1.5,3) {\small\(123\)}; 

    \node[shape=circle,draw=black] (22) at (1.5,3) {\small \(3456\)}; 

    \node[shape=circle,draw=black] (31) at (0,4.5) {\small \(E \)};

     \path [-] (00) edge (12);

    \path [-] (00) edge node[left, yshift=-5pt] {$\rho = 2$} (21);

    \path [-] (12) edge (22);

    \path [-] (21) edge node[left, yshift=5pt] {$\eta = 2$} (31);

    \path [-] (22) edge (31);

\end{tikzpicture}

}

\caption{Lattice of cyclic flats of $M_{C}$.}

\label{fig:LCF_ExR}

\end{figure}
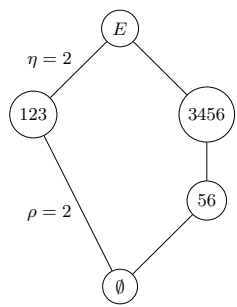

It is clear that this representation is enough to reconstruct the so-called \emph{configuration} of the matroid, \emph{i.e.}, the isomorphism type of the lattice of cyclic flats, together with the cardinality and rank of the cyclic flats. However, this data does not uniquely determine the matroid, as is shown in~\cite{eberhardt14}. 

As is proven in~\cite{westerback15}, the configuration of a representable matroid determines the minimum distance of the corresponding code, via the formula \[d_{C}=\eta(E)+1-\max_{Z\in\ZZ(M)-\{E\}}\eta(Z).\] In particular, for a binary code $C$ with $d_C>2$, all edges $Z\lessdot E$ on the ``top level'' of $\ZZ(M_C)$ must be nullity edges. Moreover, the minimum distance is then one higher than the smallest label of a top level edge in $\ZZ(M_C)$.

Recall that an $(n,k,d)$-storage code is said to have locality $(r,\delta)$ if every storage node $i\in[n]$ is contained in a set $X$ with $|X|\leq r+\delta-1$ and $d_X\geq \delta$. This is equivalent to that every node $i\in[n]$ is contained in a set $X$ with $s_X:=|X|-d_X+1\leq r$ and $d_X\geq \delta$. Now, notice that $s_X=\rho(X)+\max_{Z<_\ZZ X}\eta(Z)$, and so $s_X$ does not increase when replacing $X$ by its closure, which is a cyclic flat as $X$ is cyclic. This gives a lattice-theoretic description of $\ZZ(M_C)$, when $C$ is a binary code with $(r,\delta)$-locality. 

Surprisingly, the descriptions are qualitatively different depending on whether $\delta=2$ or $\delta>2$, \ie, whether one or more erasures can be corrected locally. This is in sharp contrast to the case when the field size is ignored, as in~\cite{westerback15}. We conclude this paper by formulating the locality criteria for binary storage codes in terms of lattices of cyclic flats. Exploiting this description to obtain quantitative bounds on the parameters $(n,k,d,r,\delta)$ is left for future research. Such bounds are also likely to suggest explicit constructions of extremal LCFs satisfying the conditions of Theorem~\ref{thm:binarylcf}.

\begin{theorem}\label{thm:binarylcf}
Let $d>2$ and let $C$ be a linear $(n,k,d,r,\delta)$-LRC over $\F_2$. Then $\ZZ=\ZZ(M_C)$ satisfies the following: 
\begin{enumerate}
\item $\emptyset$ and $[n]$ are cyclic flats.
\item Every covering relation $Z\lessdot_\ZZ[n]$ is a nullity edge labeled with a number $\geq d-1$.
\item If $\delta=2$, then for every $i\in [n]$, there is $X\in\ZZ$ with $i\in X$ such that $\rho(X)\leq r$.
\item If $\delta>2$, then for every $i\in [n]$, there is $X\in\ZZ$ with $i\in X$ such that
\begin{enumerate}[(i)]
\item Every covering relation $Y\lessdot_\ZZ X$ is a nullity edge labeled with a number $\geq \delta-1$.
\item Every cyclic flat $Y\lessdot_{\ZZ} X$ has rank $\leq r-2$
\end{enumerate} 
\end{enumerate}
\end{theorem}

\section{Conclusions and Future Work}
We have studied the lattice of cyclic flats of matroids, with a special emphasis on identifying uniform minors, and with applications to locally repairable codes over small fields. Necessary and sufficient criteria for a specified minor $M|Y/X$ to be uniform are derived in general, and in the special case of $U_4^2$-minors, necessary global criteria for $U_4^2$-avoidance are given. Finally, it is shown how these criteria dictate the structure of binary storage codes with prescribed locality parameters. Future work include translating these structural results to quantitative parameter bounds. Similar arguments are likely to be applicable when studying storage codes over other small fields, although new methods would then be needed to identify other minors than uniform ones.

\bigbreak
\textbf{Acknowledgment.}
The authors gratefully acknowledge the financial support from the Academy of Finland (grants
\#276031 and \#303819).

\bibliographystyle{splncs03}
\bibliography{references}

\end{document}